\documentclass[12pt,prd]{revtex4}%
\usepackage{graphicx}
\usepackage{amsfonts}
\usepackage{amssymb}
\usepackage{amsmath}%
\begin{document}
\title{Kottler -$\ \Lambda$\ - Kerr Spacetime}
\author{E.N. Glass and J.P. Krisch}
\affiliation{Department of Physics, University of Michigan, Ann Arbor, Michgan 48109}
\date{27 May 2004 - ver03}

\begin{abstract}
\ \newline The static Kottler metric is the Schwarzschild vacuum metric
extended to include a cosmological constant. Angular momentum is added to the
Kottler metric by using Newman and Janis' complexifying algorithm. \ The new
metric is the $\Lambda$ generalization of the Kerr spacetime. It is
stationary, axially symmetric, Petrov type \textbf{II}, and has Kerr-Schild
form.\newline PACS numbers: 04.20.-q, 04.20.Jb\ \newline\ \newline
\textbf{CONTENTS} \ \newline\ \newline$
\begin{array}
[c]{l}%
\text{I.\ INTRODUCTION}\\
\text{\rule{14pt}{0pt}A.\ Generalizations}\\
\text{\rule{14pt}{0pt}B.\ Our\ generalization}\\
\text{II.\ NJ\ APPLIED\ TO\ KOTTLER}\\
\text{III.\ MATTER\ CONTENT}\\
\text{\rule{14pt}{0pt}A.\ Energy-Momentum}\\
\text{\rule{14pt}{0pt}B.\ Angular\ momentum}\\
\text{IV.\ GEOMETRIC\ STRUCTURE}\\
\text{V.\ DISCUSSION}\\
\text{Appendix\ A:\rule{14pt}{0pt}Carter}^{\prime}\text{s\ solution}\\
\text{Appendix\ B:\rule{14pt}{0pt}Trapped\ surface}\\
\text{\rule{14pt}{0pt}\rule{14pt}{0pt}\rule{14pt}{0pt}\rule{14pt}{0pt}%
1.\ Kottler\ trapped\ surface\ location}\\
\text{\rule{14pt}{0pt}\rule{14pt}{0pt}\rule{14pt}{0pt}\rule{14pt}{0pt}%
2.\ spK\ trapped\ surface\ location}\\
\text{Appendix\ C:\rule{14pt}{0pt}Kerr-Schild\ form}%
\end{array}
$

\end{abstract}
\maketitle

\newpage

\section{INTRODUCTION}

The cosmological constant, originally included in the field equations to
create a static universe, is now included in models of the universe to drive
acceleration. How the cosmological constant effects more local scenarios, both
static and stationary, is also of interest.\ The 1918 Kottler solution,
$g_{\alpha\beta}^{Kot}$, describes matter from cosmological constant $\Lambda$
around a spherical source of mass $m_{0}$, creating a $\Lambda$ generalization
of the Schwarzschild vacuum solution. The Kottler metric is
\begin{equation}
g_{\alpha\beta}^{Kot}dx^{\alpha}dx^{\beta}=(1-2m_{0}/r-\Lambda r^{2}%
/3)du^{2}+2dudr-r^{2}(d\vartheta^{2}+\sin^{2}\vartheta d\varphi^{2}).
\label{kot-met}%
\end{equation}
When Einstein's field equations are extended to include $\Lambda
g_{\alpha\beta}$, Birkhoff's theorem extends to the Kottler solution as the
unique consequence of spherical symmetry \cite{Bon62}. Kottler reduces to the
vacuum Schwarzschild solution when $\Lambda=0$. The Kottler spacetime is an
Einstein space with Ricci tensor
\begin{equation}
R_{\alpha\beta}^{Kot}=\Lambda g_{\alpha\beta}^{Kot}.
\end{equation}

Chrusciel and Simon\ \cite{CS01} have discussed the classification of static
spherically symmetric solutions with $\Lambda$. For $m_{0}=0$ the spacetime is
either de Sitter ($\Lambda>0$) or anti-de Sitter ($\Lambda<0$). Its conformal
structure has been studied by Lake and Roeder \cite{LR77}. The Kottler
spacetime with $m_{0}\neq0$ is not cosmological since it is spherically
symmetric rather than LRS. Examining geodesic orbits shows the $\Lambda$ term
corresponds to a repulsive ($\Lambda>0$) central force of magnitude
$(\Lambda/3)r$. $\Lambda$ contributes to the periapsis precession of geodesic
orbits but does not effect gravitational lensing \cite{Lak02}. This spacetime
is Petrov type \textbf{D}, with $\Psi_{2}=-m_{0}/r^{3}$, the only non-zero
Weyl tensor component. Only the Schwarzschild mass contributes to the Weyl
tensor, which explains why $\Lambda$ has no lensing effect. With $m_{0}\neq0$,
one interpretation of $\Lambda$ in the field equations $G_{\alpha\beta}%
^{Kot}=-8\pi T_{\alpha\beta}^{Kot}=-\Lambda g_{\alpha\beta}^{Kot}$ is a vacuum
energy-momentum. In the Newtonian limit $\Lambda$ is recognized as a uniform
mass density
\begin{equation}
\rho_{\Lambda}=-\Lambda/4\pi. \label{newt-dens}%
\end{equation}
An upper limit \cite{Coh98} for $\Lambda\simeq2\times10^{-35}\ \sec^{-2}$ is
equivalent to a vacuum density of $\rho_{\Lambda}\simeq2\times10^{-45}%
\ $gm/cm$^{3}$. We will focus on $\Lambda$ as a backgound atmosphere.

With $m_{0}=0$ there is a trapped surface (horizon) at $R_{h}=\sqrt{3/\Lambda
}.$ With mass, the presence of the $\Lambda-$ atmosphere modifies the horizon
structure even in the static case, the positions being given by the roots of
the cubic function $1-2m_{0}/r-\Lambda r^{2}/3.$ While horizons in both
Schwarzschild and Kottler have constant radial coordinates, the Kottler
spacetime has two horizons for the mass range of the observed black holes.
Using $\Lambda\sim10^{-51}\ m^{-2}$ \cite{Car04} these are black holes with
$m_{0}<$ $10^{25}\ m$. One of the horizons occurs in the vicinity of the
Schwarzschild horizon and the other at larger $r$. As an example, consider the
black hole in $M87$ containing about a billion suns. For a black hole of this
size, the Kottler horizons will occur at $r\sim2m_{0}$ and $r\sim R_{h}%
\sim5.5\times10^{26}\ m.$ The radius of this horizon is\ bigger than a $13$
billion year old universe expanding at the speed of light.\ As angular
momentum is added, the horizon structure will change.

\subsection{Generalizations}

The Newman-Janis (NJ) transformation has been used recently \cite{Kim99} to
obtain the AdS$_{3}$ rotating black hole solution from the non-rotating black
hole solution of Ba\~{n}ados, Teitelboim, and Zanelli. Drake and Turolla
\cite{DT97} have used the NJ transformation to generate interior solutions for
the Kerr spacetime, focusing on bounded interiors that cannot be written in
Kerr-Schild form. Yazadjiev \cite{Yaz99} has applied the transformation to the
rotating dilaton--axion black hole.

Carter \cite{Car73} discovered a $\Lambda-$ generization of the vacuum Kerr
solution which preserves the Einstein space property. Carter's solution has
been studied by Stuchl\'{\i}k et al \cite{SBO+98}. Demianski \cite{Dem72}
found a $\Lambda-$ generalization of the NUT \cite{NCC+65} solution by
considering a complex coordinate transformation
\begin{align*}
r^{\prime}  &  =r+iF(\theta,\phi)\\
u^{\prime}  &  =u+iG(\theta,\phi)
\end{align*}
of metric
\[
ds^{2}=M(r)du^{2}+2dudr-r^{2}d\Omega^{2}%
\]
with functions $F(\theta,\phi),$ $G(\theta,\phi),$\ $M(r)$ determined from the
field equations with only $\Lambda g_{\alpha\beta}$ matter content. Vaidya
\cite{Vad76} discovered a vacuum generalization of the Kerr solution which he
called "the associated Kerr metric". Vaidya's solution is related \cite{Kra97}
to Demianski's. The complex NJ transformation \cite{NJ65},\cite{NCC+65}%
,\cite{New73} generates the vacuum Kerr spacetime from the vacuum
Schwarzschild solution. It is a special case of Demianski's transformation
with $F=-G=\cos\theta$.

An NJ transformation of the curvature tensor was done by Quevedo \cite{Que92},
producing curvature tensors which satisfy the Einstein-Maxwell equations with
$\Lambda$.

\subsection{Our generalization}

In this work we discuss the NJ spinning generalization (which we believe is a
new solution) of the static Kottler spacetime, its matter content, and trapped
surface position. The transformation can be used to investigate how $\Lambda$
will affect stationary generalizations of the Kottler spacetime with no
assumptions about the resulting matter content. The matter content will follow
directly from the field equations.

\section{NJ APPLIED TO KOTTLER}

A basis set of null vectors for $g_{\alpha\beta}^{Kot}$ is
\begin{align}
l^{\alpha}\partial_{\alpha}  &  =\partial_{r},\label{kot-tet}\\
n^{\alpha}\partial_{\alpha}  &  =\partial_{u}-(1/2)(1-2m_{0}/r-\Lambda
r^{2}/3)\partial_{r},\nonumber\\
m^{\alpha}\partial_{\alpha}  &  =(1/\sqrt{2})(1/r)[\partial_{\vartheta
}+(i/\sin\vartheta)\partial_{\varphi}].\nonumber
\end{align}
The complex coordinate transformation of the NJ algorithm is
\begin{align}
u^{\prime}  &  =u-ia\cos\vartheta,\\
r^{\prime}  &  =r+ia\cos\vartheta,\nonumber\\
\vartheta^{\prime}  &  =\vartheta,\text{ \ }\varphi^{\prime}=\varphi.\nonumber
\end{align}
Note that there are no arbitrary functions in the transformation, only the
parameter '$a$'. The transformed null tetrad is (with the primes omitted from
the new coordinates)
\begin{subequations}
\label{newkot-tet}%
\begin{align}
L^{\alpha}\partial_{\alpha}  &  =\partial_{r},\\
N^{\alpha}\partial_{\alpha}  &  =\partial_{u}-(1/2)(1-2m_{0}r/\Sigma
-\Lambda\Sigma/3)\partial_{r},\\
M^{\alpha}\partial_{\alpha}  &  =\frac{1}{\sqrt{2}(r+ia\cos\vartheta)}\left[
\partial_{\vartheta}+\frac{i}{\text{sin}\vartheta}\partial_{\varphi}%
+ia\sin\vartheta(\partial_{u}-\partial_{r})\right]  ,\\
\bar{M}^{\alpha}\partial_{\alpha}  &  =\frac{1}{\sqrt{2}(r-ia\cos\vartheta
)}\left[  \partial_{\vartheta}-\frac{i}{\text{sin}\vartheta}\partial_{\varphi
}-ia\sin\vartheta(\partial_{u}-\partial_{r})\right]  .
\end{align}
Here $\Sigma=r^{2}+a^{2}\cos^{2}\vartheta$. This transformed tetrad yields a
new metric
\end{subequations}
\begin{equation}
g_{spK}^{\alpha\beta}=L^{\alpha}N^{\beta}+N^{\alpha}L^{\beta}-M^{\alpha}%
\bar{M}^{\beta}-\bar{M}^{\alpha}M^{\beta}%
\end{equation}
or, with $\psi_{\Lambda}=1-2m_{0}r/\Sigma-(\Lambda/3)\Sigma$, and $R_{\Lambda
}^{2}=\Sigma+(2-\psi_{\Lambda})a^{2}\sin^{2}\vartheta$
\begin{align}
g_{\alpha\beta}^{spK}dx^{\alpha}dx^{\beta}  &  =\psi_{\Lambda}du^{2}%
+2dudr+(1-\psi_{\Lambda})a\sin^{2}\vartheta\,2dud\varphi\label{spk-dn-met}\\
&  -a\sin^{2}\vartheta\,2drd\varphi-\Sigma d\vartheta^{2}-R_{\Lambda}^{2}%
\sin^{2}\vartheta d\varphi^{2}.\nonumber
\end{align}
The new spacetime is stationary and axially symmetric with Killing vectors
$\partial_{u}$ and $\partial_{\varphi}$.

Since the Kerr metric has form [$\psi=1-(2m_{0}r)/\Sigma$]
\begin{align*}
g_{\alpha\beta}^{Kerr}dx^{\alpha}dx^{\beta}  &  =\psi du^{2}+2dudr+(1-\psi
)a\text{sin}^{2}\vartheta\ 2dud\varphi\\
&  -a\text{sin}^{2}\vartheta\ 2drd\varphi-\Sigma d\vartheta^{2}-[\Sigma
+(2-\psi)a^{2}\sin^{2}\vartheta]\sin^{2}\vartheta d\varphi^{2}%
\end{align*}
spinning Kottler is related to Kerr by
\begin{equation}
g_{\alpha\beta}^{spK}=g_{\alpha\beta}^{Kerr}-(\Lambda/3)\Sigma\,L_{\alpha
}L_{\beta}%
\end{equation}

\section{MATTER\ CONTENT}

\subsection{Energy-momentum}

The matter content of the spinning Kottler spacetime is abstracted from the
transformed Einstein tensor with no assumptions. The Einstein tensor has
components
\begin{equation}
G_{\alpha\beta}^{spK}=-4\Phi_{11}(L_{\alpha}N_{\beta}+N_{\alpha}L_{\beta
})-2\Phi_{22}L_{\alpha}L_{\beta}-\Lambda g_{\alpha\beta}^{spK}.
\label{spk-ein2}%
\end{equation}
with%
\begin{align}
\Phi_{11}  &  =-(\Lambda/3)a^{2}\cos^{2}\vartheta/\Sigma,\label{ein2-comp}\\
\Phi_{22}  &  =-(\Lambda/3)a^{2}(1-3\cos^{2}\vartheta)/\Sigma.\nonumber
\end{align}
For $G_{\mu\nu}=-8\pi T_{\mu\nu}$, the energy-momentum content may be
interpreted as three fluids: the first fluid (with $\Lambda g^{spK}$) can be
described as an isotropic cloud of strings, the second is a two-dimensional
rotating string fluid with density $\rho=4\Phi_{11}/8\pi$ and pressure
$p=-4\Phi_{11}/8\pi$, $\rho+p=0$. The third is a null fluid with magnitude
$2\Phi_{22}/8\pi$. The matter content imposed by Demianski is only one of the
three fluids that appear in $g^{spK}$. This particular decomposition is, of
course, not unique. Using the techniques of Coley and\ Tupper \cite{CT83} one
could describe the fluid in a number of formally equivalent ways.

\subsection{Angular momentum}

The Komar superpotential for metric $g^{spK}$ with axial Killing vector
$k^{\alpha}\partial_{\alpha}=\partial_{\varphi}$ is
\begin{equation}
U^{\alpha\beta}=(-g)^{1/2}k^{[\alpha;\beta]} \label{kom-super}%
\end{equation}
where $(-g)^{1/2}=\Sigma\sin\vartheta$. Since $g^{spK}$ is stationary, and the
rotation axis is a regular line in the spacetime, one can integrate the Komar
superpotential on a two-surface of constant '$u$' and '$r$', which bounds any
$u=const$ three-volume ($3V$). The angular momentum is given by
\begin{equation}
J^{spK}=%
{\displaystyle\oint\limits_{\partial(3V)}}
U^{\alpha\beta}(\partial_{\varphi})\,u_{[,\alpha}r_{,\beta]}d\vartheta
d\varphi.
\end{equation}
We find
\begin{equation}
J^{spK}=-16\pi ma+16\pi(\frac{\Lambda}{3})\frac{4a^{3}r}{15}.
\label{spk-angmom}%
\end{equation}
The standard Kerr angular momentum remains when $\Lambda\rightarrow0$.

\section{GEOMETRIC STRUCTURE}

Geometrically, the spinning Kottler spacetime is algebraically special, Petrov
type \textbf{II}, and the rotation axis $[\theta:0,2\pi]$ is a regular line in
the spacetime. The relation of the spK spacetime to Kerr is reflected in its
generalized Kerr-Schild form (see Appendix C)%
\begin{equation}
g_{\mu\nu}^{spK}=g_{\mu\nu}^{Kerr}-(\Lambda/3)\Sigma\,L_{\mu}L_{\nu}
\label{spk-kerr-ks}%
\end{equation}
The Ricci tensor is
\begin{equation}
R_{\alpha\beta}^{\text{spK}}=-2\Phi_{11}(L_{\alpha}N_{\beta}+N_{\alpha
}L_{\beta}+M_{\alpha}\bar{M}_{\beta}+\bar{M}_{\alpha}M_{\beta})-2\Phi
_{22}L_{\alpha}L_{\beta}+(\mathcal{R}/4)g_{\alpha\beta}^{spK}
\label{spk-ricci}%
\end{equation}
with $\Phi_{11}$ and $\Phi_{22}$ given in Eq.(\ref{ein2-comp}). The Ricci
scalar $\mathcal{R}$ (not to be confused with $R=r-ia\cos\vartheta$) is
\[
\mathcal{R}/4=(\Lambda/3)(3r^{2}+a^{2}\cos^{2}\vartheta)/\Sigma.
\]
The non-zero Weyl tensor components are%
\begin{align}
\Psi_{2}  &  =(\Lambda/3)\frac{2a^{2}\cos^{2}\vartheta}{3\Sigma}-\frac{m_{0}%
}{R^{3}},\label{psi2_spk}\\
\Psi_{3}  &  =-\frac{a\sin\vartheta}{\sqrt{2}R}\left[  (\Lambda/3)\frac
{2a\cos\vartheta}{R}+i\frac{3m_{0}}{R^{3}}\right]  ,\label{psi3_spk}\\
\Psi_{4}  &  =\frac{a^{2}\sin^{2}\vartheta}{R^{2}}\left[  -(\Lambda
/3)\frac{\Sigma}{R^{2}}+\frac{3m_{0}}{R^{3}}\right]  . \label{psi4_spk}%
\end{align}
The Kretschmann invariant is
\begin{align}
R_{\text{spK}}^{\alpha\beta\mu\nu}R_{\alpha\beta\mu\nu}^{\text{spK}}  &
=\frac{8}{\Sigma^{6}}[(\Lambda/3)^{2}\Sigma^{4}(3\Sigma^{2}+a^{4}\cos
^{4}\vartheta-4r^{2}a^{2}\cos^{2}\vartheta)\label{riemsq-spk}\\
&  +8(\Lambda/3)m_{0}r\Sigma^{2}a^{2}\cos^{2}\vartheta(3a^{2}\cos^{2}%
\vartheta-r^{2})\nonumber\\
&  +6m_{0}^{2}(\Sigma-2a^{2}\cos^{2}\vartheta)(\Sigma^{2}-16r^{2}a^{2}\cos
^{2}\vartheta)].\nonumber
\end{align}
The limit $\Lambda\rightarrow0$ yields the Kretschmann scalar for the vacuum
Kerr solution. The quadratic Ricci scalar is
\begin{equation}
R_{\text{spK}}^{\alpha\beta}R_{\alpha\beta}^{\text{spK}}=4\Lambda^{2}\left[
1-\frac{4a^{2}\cos^{2}\vartheta}{9\Sigma^{2}}(3r^{2}+a^{2}\cos^{2}%
\vartheta)\right]  . \label{riccisq-spk}%
\end{equation}
This scalar has directionality, unlike Carter's in Eq.(\ref{cart-ric-sq}).

The ergosphere and horizon structure of the spinning Kottler and Kerr
spacetimes are quite interesting to compare. To locate an horizon we search
for trapped surfaces. Using the method of Senovilla \cite{Sen02} (see Appendix
B) to identify the location of the marginally trapped surface, whose time
history is the apparent horizon, one finds
\begin{equation}
(\Lambda/3)(r^{2}+a^{2}\cos^{2}\vartheta)^{2}-\Delta=0 \label{lambda-delta}%
\end{equation}
where $\Delta=r^{2}-2m_{0}r+a^{2}$. The solutions to this equation provide an
interesting contrast to the Kerr horizon. For Kerr ($\Lambda=0$) this equation
has solutions for $a\leq m_{0}$. When $a=m_{0}$ there is a single horizon
position, $r=m_{0}$. With the $\Lambda$ field added, the extreme angular
momentum value changes and the value of '$a$' can increase. For Kerr, the
horizon position are the points were $\Delta$ cuts the $r$ axis.\ For spinning
Kottler, the positions are the multiple intersections of $\Delta$ with the
first term in equation (\ref{lambda-delta}). The presence of the cosmological
constant changes the structure of the extreme rotating solution. The
ergosurfaces bound the trapped surfaces and are found at
\begin{equation}
(\Lambda/3)(r^{2}+a^{2}\cos^{2}\vartheta)^{2}-r^{2}+2m_{0}r-a^{2}\cos
^{2}\vartheta=0. \label{spk-ergo}%
\end{equation}
The outer trapped surface touches the outer ergosurface at $\vartheta=0$. At
$\vartheta=\pi/2$, the spK ergosurface has the same equation and roots as the
Kottler trapped surface
\[
g_{(u)(u)}^{Kot}=1-2m_{0}/r-(\Lambda/3)r^{2}=0.
\]

In the vacuum Kerr solution outside a black hole, specific angular momentum
$(J/m_{0})_{\text{kerr}}$ is restricted by $m_{0}\geq a$. This restriction
preserves the existence of a trapped surface inside the ergosphere. For values
of '$a$' beyond the extreme black hole limit, a naked singularity is present
in the Kerr solution. This limit appears to change when the cosmological
constant is included, and may possess implications for microscopic black holes.

\section{DISCUSSION}

In this paper we have described the NJ-$\Lambda$ generalization of Kerr
spacetime. The matter content is spinning and reduces to the Kerr vacuum in
the $\Lambda=0$ limit. The spinning Kottler (spK) metric is similar to the
Kerr-Newman (KN) metric, and differs from KN only in the functional form of
$\psi$
\begin{align}
\text{KN}\text{: \ }  &  \psi_{Q}=1-2m_{0}r/\Sigma+Q^{2}/\Sigma
\label{psi-func}\\
\text{spK}\text{: \ }  &  \psi_{\Lambda}=1-2m_{0}r/\Sigma-(\Lambda
/3)\Sigma.\nonumber
\end{align}
The spK metric differs from Demianski's NUT generalization in several ways.
Demianski's solution (and the NUT solution) has a wire singularity on the
rotation axis. The axis is not singular in the spK solution. The spK metric
reduces to Kerr in the $\Lambda=0$ limit while Demianski's solution goes to
NUT in this limit. The NJ spin up of Kottler is also different from Carter's
$\Lambda$ generalization of Kerr. Carter's spacetime contains only a
"$\Lambda$ fluid", $\Lambda g^{\text{Cart}}$, while $g^{spK}$ has more complex
fluid content. That the two spacetimes are different is most clearly seen by
comparing the quadratic Ricci invariants, Eq.(\ref{riccisq-spk}) and
Eq.(\ref{cart-ric-sq}).

Some interesting questions remain to be examined. Future work could include a
stability analysis and an examination of geodesic orbits for this metric.
Spinning Kottler presents a new equilibrium spacetime for future study.

\appendix 

\section{Carter's solution}

In Boyer-Lindquist coordinates, Carter's metric \cite{Car73} is%
\begin{align*}
g_{\alpha\beta}^{\text{Cart}}dx^{\alpha}dx^{\beta}  &  =\frac{1}{I^{2}%
}[1-\frac{\Lambda}{3}(r^{2}+a^{2}\sin^{2}\vartheta)-\frac{2mr}{\Sigma}%
]dt^{2}-\frac{\Sigma}{\Delta_{r}}dr^{2}\\
&  +\frac{a\text{sin}^{2}\vartheta}{I^{2}}[\frac{\Lambda}{3}(r^{2}%
+a^{2})+\frac{2mr}{\Sigma}]2dtd\varphi_{_{_{B-L}}}-\frac{\Sigma}%
{\Delta_{\vartheta}}d\vartheta^{2}\\
&  -\frac{\text{sin}^{2}\vartheta}{I^{2}}[\Sigma+\frac{\Lambda}{3}a^{2}%
(r^{2}+a^{2})+a^{2}\sin^{2}\vartheta(1+2mr/\Sigma)]d\varphi_{_{_{B-L}}}^{2}%
\end{align*}
where%
\begin{align*}
\Delta_{r}  &  =(1-\frac{\Lambda}{3}r^{2})(r^{2}+a^{2})-2mr,\text{
\ \ \ }\Delta_{\vartheta}=(1+\frac{\Lambda}{3}a^{2}\cos^{2}\vartheta),\\
I  &  =1+\frac{\Lambda}{3}a^{2},\text{ \ \ \ \ \ \ \ }\Sigma=r^{2}+a^{2}%
\cos^{2}\vartheta\text{ }%
\end{align*}
(In the Black Holes article, Carter has $-\Lambda$ of $\Lambda$ here). The
limit $\Lambda\rightarrow0$ gives the Kerr solution in Boyer-Lindquist
coordinates. The limit $a\rightarrow0$ gives the Kottler solution. The Ricci
tensor is%
\begin{equation}
R_{\alpha\beta}^{\text{Cart}}=\Lambda g_{\alpha\beta}^{\text{Cart}}
\label{cart-ric}%
\end{equation}
with invariant square%
\begin{equation}
R_{\alpha\beta}^{\text{Cart}}R_{\text{Cart}}^{\alpha\beta}=4\Lambda^{2}
\label{cart-ric-sq}%
\end{equation}
Carter's metric can be mapped to ($u,r,\vartheta,\varphi$) coordinates by%
\[
dt=du+\frac{I(r^{2}+a^{2})}{\Delta_{r}}dr,\ \ \ \ d\varphi_{_{B-L}}%
=d\varphi+\frac{Ia}{\Delta_{r}}dr
\]%
\begin{align*}
g_{\alpha\beta}^{\text{Cart}}dx^{\alpha}dx^{\beta}  &  =\frac{1}{I^{2}}\left[
1-\frac{\Lambda}{3}(r^{2}+a^{2}\sin^{2}\vartheta)-\frac{2mr}{\Sigma}\right]
du^{2}+\frac{1}{I}2dudr\\
&  +\frac{2a\text{sin}^{2}\vartheta}{I^{2}}\left[  \frac{\Lambda}{3}%
(r^{2}+a^{2})+\frac{2mr}{\Sigma}\right]  dud\varphi\\
&  -\frac{2a\text{sin}^{2}\vartheta}{I\Delta_{r}}\left[  \frac{\Lambda}%
{3}(r^{2}+a^{2})(1-r^{2})-2mr\right]  drd\varphi-\frac{\Sigma}{\Delta
_{\vartheta}}d\vartheta^{2}\\
&  -\frac{\text{sin}^{2}\vartheta}{I^{{}}}\left[  \Sigma+\frac{\Lambda}%
{3}a^{2}(r^{2}+a^{2})+a^{2}\sin^{2}\vartheta(1+2mr/\Sigma)\right]
d\varphi^{2}%
\end{align*}

\section{TRAPPED SURFACE}

We follow the notion of a trapped surface given in Hawking and Ellis
\cite{HE73}.\newline For the spacetime pair of manifold $\mathcal{M}$ and
metric $g$, let $\mathcal{S}$ be a two-dimensional surface with intrinsic
coordinates $\{x^{A}:$ $A,B=2,3\}$ imbedded in $\mathcal{M}$ by parametric
equations $x^{\alpha}=F^{\alpha}(x^{A})$. Tangent vectors of $\mathcal{S}$
are
\[
\vec{e}_{A}:=\left[  \frac{\partial F^{\alpha}}{\partial x^{A}}\frac{\partial
}{\partial x^{\alpha}}\right]  _{\mathcal{S}}%
\]
with first fundamental form
\[
\gamma_{AB}=\left[  \frac{\partial F^{\mu}}{\partial x^{A}}\frac{\partial
F^{\nu}}{\partial x^{B}}g_{\mu\nu}\right]  _{\mathcal{S}}.
\]
Assume $\gamma_{AB}$ is negative definite so that $\mathcal{S}$ is spacelike.
There exist two independent null one-forms $\omega_{\mu}^{(\pm)}$ on
$\mathcal{S}$ such that%
\[
\omega_{\mu}^{(\pm)}e_{A}^{\mu}=0,\text{ \ \ \ }\omega_{\mu}^{+}\omega^{+\mu
}=0,\text{ \ \ \ }\omega_{\mu}^{-}\omega^{-\mu}=0,\text{ \ \ \ }\omega_{\mu
}^{+}\omega^{-\mu}=1,
\]
with scale freedom%
\[
\omega_{\mu}^{+}\rightarrow\tilde{\omega}_{\mu}^{+}=f^{2}\omega_{\mu}%
^{+},\text{ \ \ \ }\omega^{-\mu}\rightarrow\tilde{\omega}^{-\mu}=f^{-2}%
\omega^{-\mu}%
\]
for positive function $f^{2}$. The two second fundamental forms on
$\mathcal{S}$ are
\[
K_{AB}^{(\pm)}=-\omega_{\mu}^{(\pm)}e_{A}^{\nu}\nabla_{\nu}e_{B}^{\mu}\text{
\ with traces \ }K^{(\pm)}:=\gamma^{AB}K_{AB}^{(\pm)}%
\]
where $\gamma^{AC}\gamma_{CB}=\delta_{\ B}^{A}$. The traces allow a "trapping
scalar" $\kappa$ to be defined.%
\begin{equation}
\kappa:=2K^{(+)}K^{(-)}. \label{ts-kappa}%
\end{equation}
The traces are the expansions of the two null geodesic congruences emerging
orthogonally from $\mathcal{S}$. Thus, $\mathcal{S}$ is trapped if both
congruences converge.

\begin{center}
$\mathcal{S}$ is $\left[
\begin{array}
[c]{c}%
\text{trapped}\\
\text{marginally trapped}\\
\text{non-trapped}%
\end{array}
\right]  $ if $\left[
\begin{array}
[c]{c}%
\kappa>0\\
\kappa=0\\
\kappa<0
\end{array}
\right]  $.
\end{center}

\subsection{KOTTLER TRAPPED SURFACE LOCATION}

First consider the static Kottler trapped surface:
\begin{equation}
\partial_{u}\cdot\partial_{u}=1-2m_{0}/r-(\Lambda/3)r^{2}=0. \label{kot-trap}%
\end{equation}
For $\Lambda<0$ there is one real root $r_{0}$ such that $0<r_{0}<2m_{0}$.
\ With $\Lambda>0$ there are three cases.

(1) For $3m_{0}>\Lambda^{-1/2}$ there is one real root $<0$.

(2) For $3m_{0}=\Lambda^{-1/2}$ there are three real roots
\[
r_{1}=r_{2}=\Lambda^{-1/2}>0,\text{ \ \ }r_{3}=-2\Lambda^{-1/2}.
\]

(3) For $3m_{0}<\Lambda^{-1/2}$ there are three distinct roots ($r_{4}$,
$r_{5}$, $r_{6}$)
\[
0<2m_{0}<r_{4}<3m_{0}<r_{5},\text{ \ \ \ }r_{6}<0.
\]

\subsection{spK TRAPPED SURFACE LOCATION}

Since the roots of a quartic locate the spK trapped surfaces, we use
\textit{Descartes' Rule of Signs} \cite{Wil62} to enumerate the roots: The
number of positive roots of equation $f(x)=0$ with real coefficients is equal
to the number of sign changes in the polynomial $f(x)$ or is less than that
number by a positive integer. The number of negative roots is equal to the
number of sign changes in $f(-x)$ or is less than that number by a positive
integer. A root of multiplicity $n$ is counted as $n$ roots.

We examine the quartic $\Delta_{\Lambda}=0$ at $\vartheta=0$.
\begin{equation}
(\frac{\Lambda}{3})r_{H}^{4}+(2\frac{\Lambda}{3}a^{2}-1)r_{H}^{2}+2m_{0}%
r_{H}+(\frac{\Lambda}{3}a^{2}-1)a^{2}=0
\end{equation}
($\Lambda>0$)\newline\ \ \ \ (i) $\Lambda>3/a^{2}$,\ \ \ \ \ \ \ \ polynomial
signs \ $[+$ $\ +$ $\ +$ $\ +]$. \ No sign changes $\Rightarrow$ no
roots.\newline\ \ \ \ (ii) $\Lambda>3/(2a^{2})$,\ \ \ \ polynomial signs
\ $[+$ $\ +$ $\ + $ $\ -]$. \ One sign change $\Rightarrow$ one real root.\ 

For $\Lambda=3/a^{2}$ the one real root is%
\begin{equation}
r_{H}=(m_{0}a^{2})^{1/3}b_{0}^{1/3}-(a^{2}/3)(m_{0}a^{2})^{-1/3}b_{0}%
^{-1/3},\text{ \ \ \ }b_{0}=\sqrt{1+a^{2}/(3m_{0}^{2})}-1.
\end{equation}
($\Lambda<0$)\ \ \ \ \ polynomial signs \ $[-$ $\ -$ $\ +$ $\ -]$. \ Two sign
changes $\Rightarrow$ two real roots or no roots.\newline

The quartic $\Delta_{\Lambda}=0$ at $\vartheta=\pi/2$ is
\begin{equation}
(\Lambda/3)r_{H}^{4}-r_{H}^{2}+2m_{0}r_{H}-a^{2}=0.
\end{equation}
($\Lambda>0$)\newline\ \ \ \ (i) $f(r_{H})$,\ polynomial signs \ $[+$ $\ -$
$\ +$ $\ -]$. \ Three sign changes $\Rightarrow$ three real roots or one
root\newline\ \ \ \ (ii) $f(-r_{H})$,\ polynomial signs \ $[+$ $\ -$ $\ -$
$\ -]$. \ One sign change $\Rightarrow$ one real root.\ \newline For
$\Lambda=1/(4a^{2})$ the four roots are%
\begin{align}
r_{H1,H2}  &  =m_{0}(h_{2}\pm h_{3}),\\
r_{H3,H4}  &  =m_{0}(-h_{2}\pm h_{3})\nonumber
\end{align}
with
\begin{align*}
h_{1}  &  =[(1/3)(a/m_{0})^{4}-(8/27)(a/m_{0})^{6}]^{1/3},\\
h_{2}  &  =[3h_{1}+2(a/m_{0})^{2}]^{1/2},\\
h_{3}  &  =[4(a/m_{0})^{2}-3h_{1}-6(a/m_{0})^{2}/h_{2}]^{1/2}.
\end{align*}

($\Lambda<0$)\ \ \ \ \ polynomial signs \ $[-$ $\ -$ $\ +$ $\ -]$. \ Two sign
changes $\Rightarrow$ two real roots or no roots.

The differences are clearly seen in the graph below, a plot of $(x-1)^{2}$ and
$(\Lambda m_{0}^{2}/3)x^{4}$ versus $x$ ($x=r/m_{0}$), $\Lambda m_{0}%
^{2}/3=0.1.$ With no cosmological constant, the horizon positions would be the
intersection of the decreasing graph with the $x$ axis at $x=1,$ $r=m_{0}%
$.\ With a cosmological constant, the trapped surface positions are the
multiple intersections of the two curves. For the parameter chosen, the four
intersections occur at $x=-6.34,0.86,1.32$ and $4.16$. Only two of the
intersections are shown in the graph.As $\Lambda$ decreases, the outermost
trapped surface moves to larger values. For $\Lambda m_{0}^{2}/3=.001,$ the
positive~intersections~are~$x~=~.98,~1.02,~53.8.~$

\begin{center}
$%
{\includegraphics[
height=2.5356in,
width=2.5668in
]%
{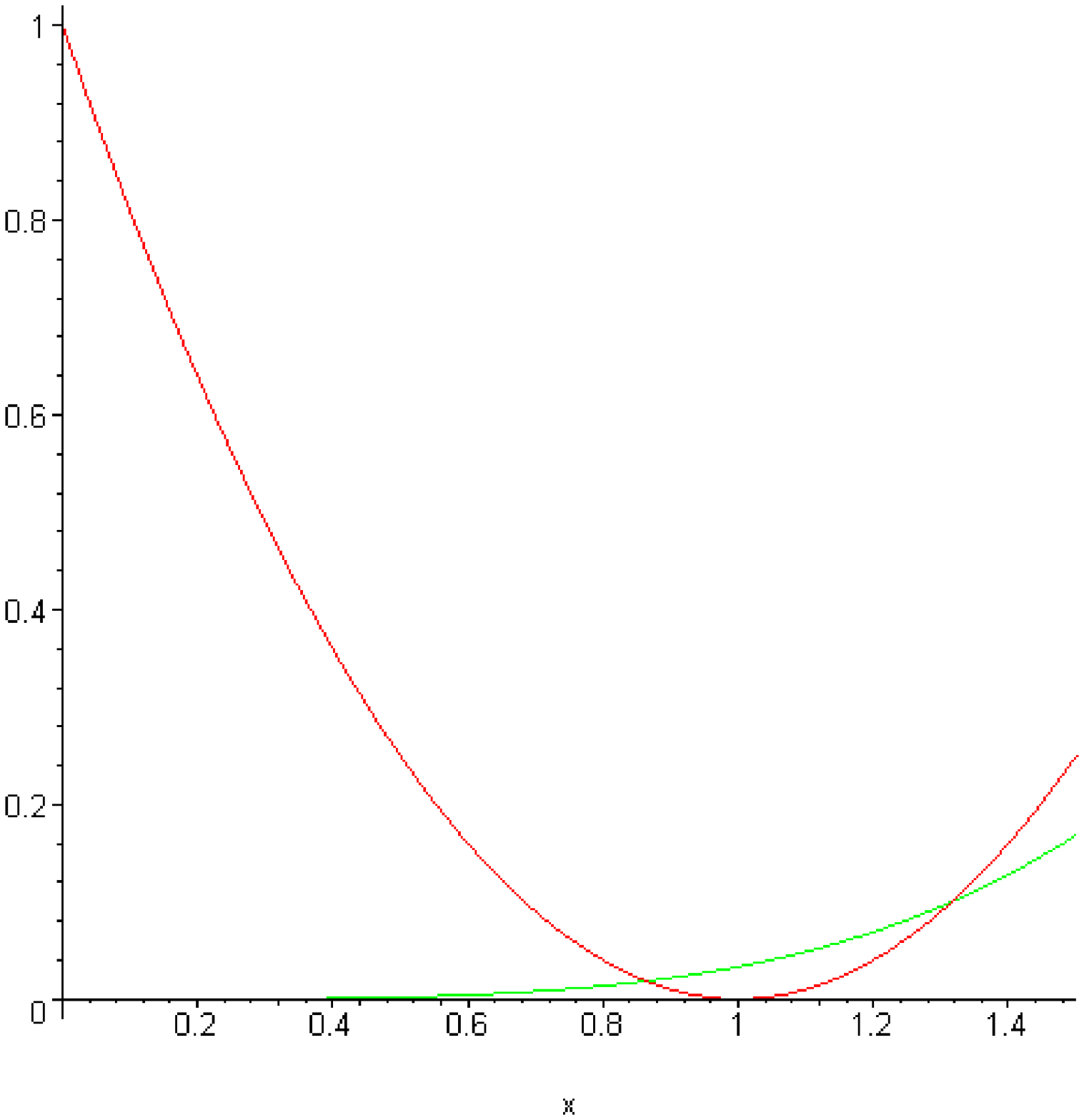}%
}
$

Figure 1. Trapped surface positions
\end{center}

\section{Kerr-Schild form}

Consider $\hat{g}_{\alpha\beta}=g_{\alpha\beta}+H(x^{\mu})l_{\alpha}l_{\beta}%
$, where $l^{\alpha}$ is null with respect to both metrics, $l^{\alpha
}l^{\beta}g_{\alpha\beta}=l^{\alpha}l^{\beta}\hat{g}_{\alpha\beta}=0$.
Xanthopoulos' \cite{MX89} generalized superposition theorem states:
\begin{equation}
\text{If }G_{\mu\nu}(g)=0\text{ and }D_{g}G_{\mu\nu}(H\,ll)=0\text{, then
}G_{\mu\nu}(\hat{g})=0. \label{ks-thm}%
\end{equation}
Here $D_{g}G_{\mu\nu}$ represents the first functional derivative of the
Einstein tensor, i.e. Einstein's equations linearized about $g$. The
generalized Kerr-Schild form of $g^{spK}$ is vacuum to non-vacuum. See
Sopuerta \cite{Sop98} and references therein.

\end{document}